# ETHERNET PACKET PROCESSOR FOR SoC APPLICATION

Raja Jitendra Nayaka[1], R. C. Biradar[2]

[1]Research and Development, ITI Limited. Bangalore, INDIA
[1]`rjnayaka@yahoo.com`
[2]Department of Electronics and Communication Engineering
[2]REVA Institute of Technology and Management, Bangalore, INDIA.
[2]`raj.biradar@revainstitution.org`

*ABSTRACT*

*As the demand for Internet expands significantly in numbers of users, servers, IP addresses, switches and routers, the IP based network architecture must evolve and change. The design of domain specific processors that require high performance, low power and high degree of programmability is the bottleneck in many processor based applications. This paper describes the design of ethernet packet processor for system-on-chip (SoC) which performs all core packet processing functions, including segmentation and reassembly, packetization classification, route and queue management which will speedup switching/routing performance. Our design has been configured for use with multiple projects ttargeted to a commercial configurable logic device the system is designed to support 10/100/1000 links with a speed advantage. VHDL has been used to implement and simulated the required functions in FPGA.*

## 1. INTRODUCTION.

With the advancements in networking technology, most networks converge on Ethernet and IP as the dominant transport technology and the ability to manage and process the IP packets at high speed is a key to offering successful new services. Packet processors intercept individual IP data packets and to process them using application software which can be easily enhanced to add new capabilities. High speed packet processing is critical for low-latency applications such as multimedia, Voice over IP (VoIP), and security. Increasingly, hardware accelerators for security and pattern matching are built into the multicore processors used in packet processors. Multicore packet processing building blocks are found in applications across all elements in the converged communication network and enable the convergence of multiple functions and diverse services from different systems into a single system.

The Ethernet hardware NIC adds its own wrapper (the Ethernet header and trailer) to each packet to direct it to the correct destination on the local network. If the packet's ultimate destination is somewhere off the local network, the Ethernet header added by the sending machine will point to a router or switch as its destination address. The router will open the packet, strip off the Ethernet wrapper, read far enough to find the ultimate destination address and re-wrap the packet, giving it a new header that will send it on the next hop of its journey. At the receiving end, the process is reversed. The packet is read by the NIC (Network Interface Card) at the receiving machine which strips off the Ethernet header and passes the packet up to the appropriate protocol stack. The protocol stack reads and strips off its headers and passes the remaining packet contents on up to the application or process to which it was addressed, reassembling the chunked data in the correct order as it arrives. The packet format is as shown in fig1

| 62 bits | Preamble used for bit synchronization |
|---|---|
| 2 bits | Start of Frame Delimiter |
| 48 bits | Destination Ethernet Address |
| 48 bits | Source Ethernet Address |
| 16 bits | Length or Type |
| 46 -1500 bytes | Data |
| 32 bits | Frame Check Sequence |

Fig1.Ethernet Frame Format

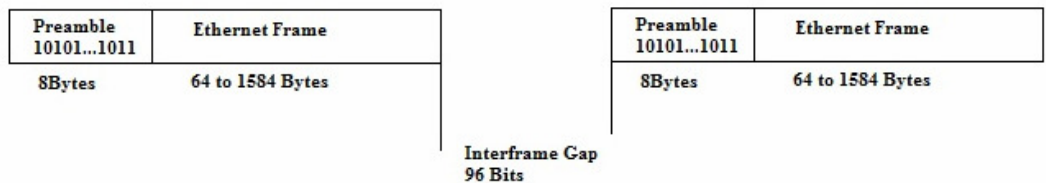

Fig 2. Decoding Ethernet Frame

Each Ethernet frame is preceded by a 64 bit preamble that consists of a bit pattern of alternating 1's and 0's, and has the last two bits set to "11" as shown in fig2. This bit pattern is used by the receiver to synchronize with the bit timing of the frame. The last two bits indicate the beginning of the Ethernet frame. Consecutive frames must be separated by a gap of 96 bits, which corresponds to 96 nanoseconds on a 100 Mbps Ethernet network.

## 1.1 SYSTEM ON CHIP (SoC)

System-on-a-chip or system on chip (SoC or SOC) refers to integrating all components of a computer or other electronic system into a single integrated circuit (chip). It may contain digital, analog, mixed-signal, and often radio-frequency functions – all on a single chip substrate. A typical application is in the area of embedded systems. The contrast with a microcontroller is one of degree. Microcontrollers typically have under 100K of RAM (often just a few KB) and are single-chip-systems; whereas the term SoC is typically used with more powerful processors, capable of running software such as Windows or Linux, which need external memory chips (flash, RAM) to be useful, and which are used with various external peripherals increasing chip integration to reduce manufacturing costs and to enable smaller systems. Many interesting systems are too complex to fit on just one chip built with a process optimized for just one of the system's tasks.

A SoC consists of both the hardware and the software that controls the microcontroller, microprocessor or DSP cores, peripherals and interfaces. The design flow for an SoC aims to develop this hardware and software in parallel. Most SoCs are developed from pre-qualified hardware blocks for the hardware elements described above, together with the software drivers that control their operation. Of particular importance are the protocol stacks that drive industry-standard interfaces like USB. The hardware blocks are put together using CAD tools; the software modules are integrated using a software development environment. A key step in the design flow is emulation: the hardware is mapped onto an emulation platform based on a field programmable gate array (FPGA) that mimics the behavior of the SoC, and the software modules are loaded into the memory of the emulation platform. Once programmed, the emulation platform enables the hardware and software of the SoC to be tested and debugged at close to its full operational speed. After emulation the hardware of the SoC follows the place and route phase of the design of an integrated circuit before it is fabricated.

### 1.2 OUR CONTRIBUTIONS

In this paper, we provide the design and implementation aspects of Ethernet packet processors for new generation IP Network technology, which is designed to address the performance and flexibility problems of new generation IP products.

## 2. LITERARURE SURVEY

General purpose processors cannot provide wire speed performance for Packet processing and analysis. Even integrating embedded processor in to the NIC cannot handle packet rates for the highest-speed networks. Embedded ASIC hardware has extremely high cost and is difficult to implement and requires many months of fabricate a chip for even small changes. The embedded processor can handle only specified functionality with wire speed performance. Packet processing needs balance between the architecture and network flow. Developing Real time packet Analysis which can receive the packets processes the packets and forwards the packet with wire speed while utilizing the maximum network bandwidth as well as to marinating security in the network is an ideal application. Network traffic analysis includes capturing of data from network and inspecting of data at each layer. The exponential growth of Internet traffic, network bandwidth and Internet based applications rise problems of performance, flexibility in network traffic analysis. Flexibility achieved through the programmable devices like General purpose processors and performance can achieve through hardwired solutions like Field Programmable Gate Arrays. With General purpose processors we can't achieve wire speed performance; with FPGA technology Network processors are new generation technology, which is designed to address the performance and flexibility problems. Network processors, analyzes Lower level layers by hardware and higher level layer by software with parallel and pipelined architecture. With multiple micro engines and with parallel and pipelining programming architecture network processors makes network processing at wire speed [18][19][20][21][22].

The features of Network Processors are compared from the following perspectives.
*Performance* - by executing key computational kernels in hardware, NPs are able to perform many applications at wire speed.
*Flexibility* – having software as a major part of the system allows network equipment to easily adapt to changing standards and applications
*Fast TTM* – designing software is much faster (and cheaper) than designing hardware of equivalent functionality
*Power* – while NPs may not be embedded in energy-sensitive devices (like handhelds), their power consumption is important for cost reasons (e.g. implications on packaging).

Major goal of the design when creating a product, which may include functionality, performance, power consumption and manufacturing cost. Functionality provides a more detailed description of what the product does, Performance means the processing speed of the product, which may be a combination of soft deadlines such as approximate time to perform a user-level function and hard deadlines by which a particular operation must be completed. Power consumption gives a rough idea of how much power the product can consume. And manufacturing cost primarily defines the cost of the hardware components. Beyond this, we must also consider some other important requirements in system design: time-to-market, design cost and quality.

*Performance and flexibility.* Traditional network processing mainly focuses on forwarding packet at high speed in order to eliminate the network bottlenecks. Network devices are expected to perform at high speed with low latency. However, as the Internet Protocol keeps maturing, newer protocols have been emerging and will emerge in the future. Such newer protocols include security, signaling, and various network managements, etc. Network devices are also expected to flexibly support these newer protocols and applications with high performance.

## 3. PROPOSED ARCHITECTURE AND DESIGN METHODOLOGY

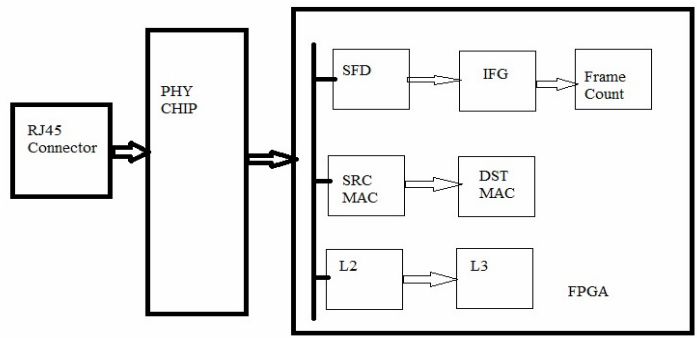

Fig: Block Diagram Of Ethernet Packet Processor

Fig 3. Ethernet Packet Processor

Those general purpose processors based architecture suffers from very low performance that hinder them from being suitable for many applications, such as L2/L3 switches, routers and Gigabit routers. On the other hand, Application Specific processors are not suitable for applications that require high performance and low power requirements, However Ethernet packet processor allow much faster packet processing. The potential solution is to make use of the Field Programmable Gate Arrays (FPGAs), have introduced a great deal of speed and flexibility and performance into machine fast controls and operations. They are extensively used to implement highly specialized tasks where simplicity, low production cost, and reliability are big assets. Even if simple and reliable, these systems need means for communications

An outline of the proposed architecture design is shown in Fig 3. Main components are Ethernet packet Processor, PHY (10/100/1000) chip and RJ45 Connector. Ethernet Packet Processor consists of five VHDL Modules. The core functionality is implemented in the Aggregate module, which has been custom-designed using VHDL. Such as detection of SFD, source MAC, destination MAC, payload length and CRC detection/calculation. A clean separation between these module and rest of the design allows for flexibility and portability, benefiting from well defined independent interfaces. Networking cores provides fertile ground for designing highly modular and re-usable components. Packet processing in FPGA is done by a chain of dedicated pipelined blocks. We implemented programmable synchronous pipeline arrays of individual block for high performance. The increase in the number of cores that can be integrated on a single chip has forced the designer to use computer network concepts for design of System on Chip (SoC).

The Ethernet Packet Processor analyzes the receiving frame. Open the packet, strip off the Ethernet wrapper, read far enough to find the required field. It identifies the type of Ethernet encapsulation, type of protocol, and extracts the fields in the packet needed by the Address look-up. The Packet processor performs start of frame preamble (SFD), interframe gap (IFG) detection, and Packet length count, source. Destination MAC address and Layer2/Layer3/Layer4 parsing to extract information from the headers of these three layers. Therefore, protocols of these three layers have to be considered. The CRC calculation is done to check integrity of transmitted and received frame.

The identification of SFD and calculation/detection of CRC is found to be time and memory consuming task in all IP based products. In this design we proposed hardware acceleration of these tasks to offload processor tasks. Similarly source and destination MAC address, IP address extraction and Frame length count is important tasks in switching and routing, EPP provides pipelined blocks to meet these task to improve performance of next generation IP products

### 3.1 PROOF OF CONCEPT AND APPLICATIONS

EPP play role hardware acceleration in design of new generation IP products for achieving high performance. The usage of EPP in design of switches and routers is discussed.

### 3.1.1 LAN Switch Design Using Ethernet Packet Processor.

LAN switches rely on packet switching. The switch establishes a connection between two segments and keeps the connection just long enough to send the current packet. Incoming packets, which are part of an Ethernet frame, save to a temporary memory area. The temporary memory area is a buffer. The switch reads the MAC address that is in the frame header and compares the address to a list of addresses in the switch lookup table. In a LAN with an Ethernet basis, an Ethernet frame contains a normal packet as the payload of the frame. The frame has a special header that includes the MAC address information for the source and destination of the packet. Traditionally switch that uses store and forward saves the entire packet to the buffer and checks the packet for Cyclic Redundancy Check (CRC) errors or other problems. If the packet has an error, the packet is discarded. Otherwise, the switch looks up the MAC address and sends the packet on to the destination node. Many switches combine the two methods by using cut-through until a certain error level is reached, then changing over to store and forward. Very few switches are strictly cut-through because this provides no error correction [2][3][4]. Since processor based switch design is busy in collecting information required for switching and switching between port lead to latency. Multiple latencies resulting from this scheme could improve the overall performance when combined with Ethernet Packet Processor in SoC as shown figure below.

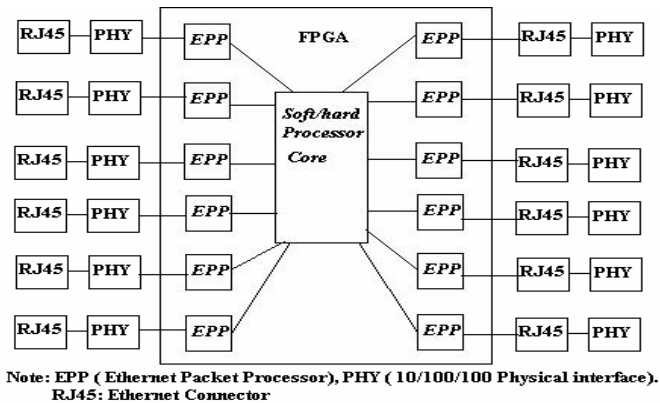

Fig 4: High capacity LAN Switch Architecture.

### 3.1.2 Router Design Using Ethernet Packet Processor.

Routers have traditionally been implemented purely in software. Because of the software implementation, the performance of a router was limited by the performance of the processor executing the protocol code. To achieve wire-speed routing, high-performance processors together with large memories were required. This translated into higher cost.

Next Generation IP Routers will use multigigabit networking technologies where IP routers will be used not only to interconnect backbone segments but also to act as points of attachments to high performance wide area links[1]. Special attention must be given to new powerful architectures for routers in order to play that demanding role. Therefore, the design of high speed IP routers has been a major area of research. Advances in optical networking technology are pushing link rates in high speed IP routers beyond 10G and 40 Gbps. Such high rates demand that packet forwarding in IP routers must be performed in hardware. The FPGA based cores are reconfigured to take into account changes in the bandwidth demands and routing characteristics [15]. While the FPGA is being reconfigured, all traffic is routed by the host software. When reconfiguration is finished, selected virtual networks are shifted back to the hardware based on their performance requirements.

When a router receives a packet, The router will open the packet, strip off the Ethernet wrapper, read far enough to find the ultimate destination address and re-wrap the packet, giving it a new header that will send it on the next hop the router looks at the Layer 3, or Network layer, source and destination addresses to determine the path for the packet to take. This activity is Layer 3 (Network) networking activity.

The evolution of IP router designs and highlights the major performance issues affecting IP routers. The need to build fast IP routers is being studied in a variety of ways. We analyzed in detail the various router mechanisms needed for high speed operation. In particular, we examine the architectural constraints imposed by the various router design alternatives. Avoiding centralized processor route lookups, and administration, is the requirements of new generation router design, router hardware can be made more reliable by adding Ethernet packet processor could be one method for fast packet processing

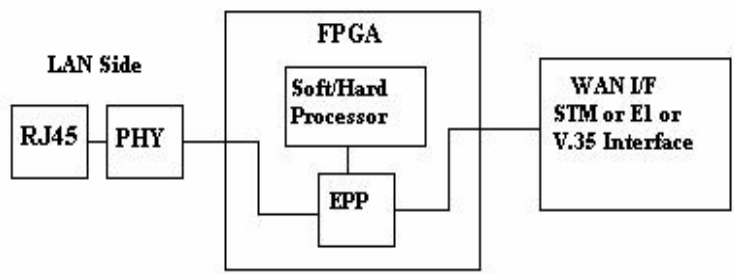

Note: EPP ( Ethernet Packet Processor), PHY ( 10/100/100 Physical interface).
RJ45: Ethernet Connector

Fig 5: High Speed high performance Router Design

## 4. CONCLUSION

A novel design methodology to design a domain specific packet processor has been introduced to have hardware acceleration for process and pass packets at high speed. We have designed an interface that directly translates the way packets need to be processed into a simple clean pipeline that has enough flexibility to allow for designing some powerful extensions to a basic switches and routers. Using this methodology, a very compact domain specific SoC core can be designed while maintaining the high speed requirements of Ethernet/IP switches and router. This Method finds wide application in design of Next generation Ethernet based products including high performance high capacity L2/L3 switches and router is possible with this method.

## REFERANCES

[1] Boosting the performance of PC-based software routers with FPGA-enhanced line cards Andrea Bianco, Robert Birke, Jorge M. Finochietto, Giulio Galante_,Marco Mellia,, Fabio Neri,, Michele Petracca Dipartimento di Elettronica, Politecnico di Torino, 10129 Torino,
[2] Z. L. A. Kennedy, X.Wang and B. Liu, "Low power architecture for high speed packet classification," in Proc. ANCS, 2008.
[3] V. Srinivasan and G. Varghese, "Fast address lookups using controlled prefix expansion," ACM Trans. Comput. Syst., vol. 17, pp. 1–40, 1999.
[4] S. Sahni and K. S. Kim, "An O(log n) dynamic router-table design," IEEE Transactions on Computers, vol. 53, no. 3, pp. 351–363, 2004.
[5] H. Lu and S. Sahni, "O(log n) dynamic router-tables for prefixes and ranges," IEEE Transactions on Computers, vol. 53, no. 10, pp. 1217–1230, 2004.
[6] K. S. Kim and S. Sahni, "Efficient construction of pipelined multibittrie router-tables," IEEE Transactions on Computers, vol. 56, no. 1,pp. 32–43, 2007.
[7] H. Le, W. Jiang, and V. K. Prasanna, "A SRAM-based architecture for trie-based IP lookup using FPGA," in Proc. FCCM '08, 2008.
[8] H. Fadishei, M. S. Zamani, and M. Sabaei, "A novel reconfigurable hardware architecture for IP address lookup," in Proc. ANCS '05, 2005, pp. 81–90.
[9] F. Baboescu, D. M. Tullsen, G. Rosu, and S. Singh, "A tree based router search engine architecture with single port memories," in Proc.ISCA '05, 2005, pp. 123–133.
[10] W. Jiang and V. K. Prasanna, "A memory-balanced linear pipeline architecture for trie-based IP lookup," in Proc. HOTI '07, 2007, pp.83–90.
[11] R. Sangireddy, N. Futamura, S. Aluru, and A. K. Somani, "Scalable, memory efficient, high-speed IP lookup algorithms," IEEE/ACM
Trans. Netw., vol. 13, no. 4, pp. 802–812, 2005.
[12] M. Blagojevic, A. Smiljanic: Design of Multicast Controller for High-capacity Internet Router, Electronic Letters, Vol. 44, No. 3, Jan. 2008, pp. 255 – 256.
[13] M. Petrović, A. Smiljanić: Design of the Scheduler for the High-capacity Non-blocking Packet Switch, IEEE Workshop on High Performance Switching and Routing, Poznan, Poland, June 2006.
[14] C. Partridge and P. P. Carvey, "A 50 Gb/s IP Router," IEEE/ACM Transactions on Networking, vol. 6, pp. 237–248, June 1998.
[15] J. R. Hess and D. C. Lee, "Implementation and Evaluation of a Prototype Reconfigurable Router," in IEEE Symposium on FPGAs for Custom Configurable Computing Machines, pp. 260–264, April 1999.
[16] The Programmable Logic Databook. Xilinx Databook Xilinx Inc., 1999.
[17] V. Kumar, et al., "Beyond Best Effort: Router Architectures for the Differentiated," IEEE Communications Magazine, vol.36, (no.5), IEEE, May 1998. p.152-64.
[18] "Packet Reordering in Network Processors" S. Govind1, R. Govindarajan1;2 and Joy . Indian Institute of Science, Bangalore 560012, India.
[19] "The Challenge for Next Generation Network Processors", Whitepaper, Agere systems, Aril2001.
[20] "On the Deployment of VoIP in Ethernet Networks: Methodology and Case Study"Khaled Salah**

[21] "10 Gbit/s Line Rate Packet Processing Using Commodity Hardware: Survey and new Proposals" Luigi Rizzo, Luca Deri, Alfredo Cardigliano ANCS'10, October 25–26, 2010, La Jolla, CA, USA

[22] "A Survey on Network Processors " Md. Ehtesamul Haque and Md. Humayun Kabir Department of Computer Science and Engineering Bangladesh University of Engineering and Technology, Dhaka 1000, Bangladesh April 3, 2007

**Athour's Biography**

**Raja Jitendra Nayaka,** Working as Senior Engineer at R&D, ITI Ltd, Govt Of India, He has over 18Yrs experience in design and development of telecom products, He has vast experience in Switching, Transmission, Internet, SDH, and optical communication, His field of interest is telecom and FPGA based designs.

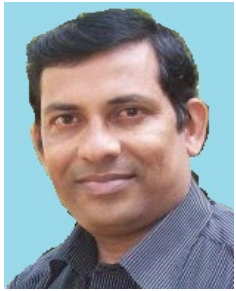

**Raja jitendra Nayaka**

**Dr. R. C. Biradar** is working as Professor in ECE Department Reva Institute of Technology and Management, Bangalore, India. He obtained his Ph. D from VTU Belgaum, India. He has many publications in reputed national/international journals and conferences. Some of the journals where his research articles published are Elsevier, IET and Springer publications having very good impact factors. His research interests include multicast routing in mobile ad hoc networks, wireless Internet, group communication in MANETs, software agent technology, network security, multimedia communication, VLSI design and FPGA, etc. He is a reviewer of various reputed journals and conferences and chaired many conferences. He is a member IETE (MIETE) India, member IE (MIE) India, member ISTE (MISTE) India and member of IEEE (USA) and member of IACSIT. He has been listed in Marqui's Who's Who in the World (2012 Edition), USA and Top 100 Engineers by IBC, UK.

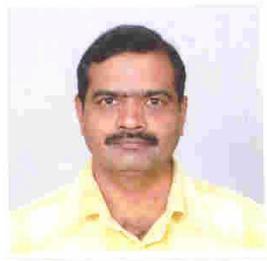

**Dr. R. C. Biradar**